\documentclass[12pt,amsfonts]{iopart}
\def\({\left(}
\def\){\right)}
\def\[{\left[}
\def\]{\right]}
\def\be{\begin{eqnarray}}
\def\ee{\end{eqnarray}}
\def\ne{\nonumber\end{eqnarray}}
\def\G{{\Gamma}}
\def\res#1{\mathop{\rm Res}\limits_{#1}}

\begin{document}
\title{Modification of Abel-Plana formula for functions with non-integrable
branch-points}
\author
{I. Fialkovsky \dag}
\address
{\dag \ High Energy Physics Department
    St Petersburg State University, Russia}

\begin{abstract}
The Abel-Plana formula is a widely used tool for calculations in
Casimir  type problems. In this note we present a particular explicit
modification of the Generalized Abel-Plana formula for the
functions with non-integrable branch-point singularities.
\end{abstract}


\section{Introduction}
Among numerous methods used for calculations in quantum field theory (QFT), an
important role is played by those exploiting analytical
properties of functions. One of such methods is the summation
formula of Abel-Plana (APF) \cite{Evgrafov} which is widely
used for calculations in Casimir problems in different
configurations \cite{Bordag}, and connected issues~\cite{Grib}.

The most frequently used form of the APF is the following \cite{Saharian 00}
\be
\sum_{n=0}^\infty
f(n)=\int_0^\infty f(x) dx
    +\frac12f(0)+i\int_0^\infty \frac{f(ix)-f(-ix)}{e^{2\pi x}-1}dx
    \label{APF1}.
\ee
It is applicable to functions satisfying the following convergence condition
\be
    \lim_{y\to\infty}e^{-2\pi y}|f(x+iy)|=0
\label{lim}
\ee
uniformly on any finite interval of $x$. This condition is
naturally meat within the framework of QFT even for formally
divergent series when (any) appropriate regularization scheme is applied.

There is another important condition of validity of (\ref{APF1}) ---
the analyticity of $f(x)$  in the right half-plane. This condition however cannot
be equally natural satisfied in field theoretical calculations and should be addressed
independently in each case. Unfortunately, in the literature (see for instance \cite{Scandurra})
it is not always payed enough (if any) attention to verification of this condition.
On the other hand, one easily see that direct application of (\ref{APF1}) to such `trivial'
functions as
\be
    f=\frac1{n^2+a^2},\quad \frac1{n^4+a^2 n^2+b^2}
\ee
leads to incorrect answers.

The major ever research on the APF and it generalizations
to different classes of functions is presented in \cite{Saharian 00}
and \cite{Saharian 07}. There is not only
a treatment of some elementary functions presented, but Bessel functions
are also considered.

However, it is not always easy to apply a cumbersome generalized formula in particular
cases, and explicit expressions are needed. In particular, for the functions possessing
non-integrable branch-point singularities on the imaginary axe an explicit form of APF is missing.
Summation of such functions appears in calculation of the Casimir energy in the
electromagnetic case with semi-transparent cylindrical shell \cite{FiMaPi}.

In this note we present a derivation of explicit summation formula for this case.

\section{APF for non-integrable branch-point singularities}
The derivation of generalized Abel-Plana formula  is based
on integration of a pair of functions along a contour in a complex plane. The contour
goes in part along the imaginary axe (for details see \cite{Saharian 07}).
Consequently any singularities
at $z=i a$, $a\in \Re$ need careful and detailed study. Integrable branch points as well
as normal poles (of arbitrary order) do not bring particular problems as they
can be expressed in terms of straightforward integrals or residues.
However, the merging of two types of singularities
must be treated independently. Such behavior of a function could be represented as
\be
    f(x)=\frac1{(x^2+q^2)^{k+1/2}},\qquad k=1,2,\ldots
    \label{f(x)}
\ee
We consider here the simplest case of a `naked' singularity as the summand
function, but further generalizations are immediate.

A direct study of the above mentioned contour integral is rather cumbersome,
and it comes out easier to start with the following form of APF
\be
\sum_{n=1}^\infty \frac1{(n^2+q^2)^k\sqrt{n^2+w^2}}
    - \int_0^\infty \frac{dx}{(x^2+q^2)^k\sqrt{x^2+w^2}}=
\label{APF4}
\ee
$$
    =-\frac{1}{2 q^{2 k}w}
    - i\pi \res{z= iq}\(\frac{g(z)}{\sqrt{x-iw}}\,\frac1{(x-iq)^k}\)
    +2(-1)^k \int_w^\infty h(x)\frac{dx}{(x-q)^k\sqrt{x-w}}
$$
which can be derived from combination of (3.22) and (3.33)
\cite{Saharian 07} and is valid for $w>q>0$.
For convenience we introduced the following notation
\be
    g(x)=\frac{-2}{(x+iq)^k\sqrt{x+iw}}\,\frac1{e^{-2\pi i x}-1}, \quad
    h(x)=-\frac{(i)^{k+1/2}}2 g(ix).
\ee

Let one consider the limit $q\to w$. The
LHS of (\ref{APF4}) is perfectly convergent in this limit. So must do the RHS
according to the analytical continuation principle.

To investigate the RHS behavior in detail, we first construct
explicitly the residue at $z=iq$. Decomposing $\frac{g(z)}{\sqrt{x-iw}}$ into a Taylor
series at this point
and exploiting an obvious connection between derivatives of $h$ and $g$
$$
    h^{(j)}(x)=-\frac{(i)^{k+j+1/2}}2 g^{(j)}(ix)
$$
we can write for the residue
\be
(-1)^{k-1} \sqrt\pi \sum_{j=0}^{k-1} h^{(j)}(q)
\frac{\G(k-j-1/2)}{\G(k-j) \G(j+1)}
        \,\frac{1}{(w-q)^{k-j-1}}
\label{Res}\ee

On the other hand, for the integral part of RHS in (\ref{APF4}) we can
construct the following decomposition
\be
I=\int_w^\infty \(h(x)-[h(x)]_q^{k-1}\)\frac{dx}{(x-q)^k\sqrt{x-w}}
\label{I}\\ \qquad
    +\int_w^\infty [h(x)]_q^{k-1}\frac{dx}{(x-q)^k\sqrt{x-w}}.
\ne
where we have subtracted the first $k$ Taylor terms
of $h(x)$ at $x=q$
\be
[h(x)]_q^{k-1}\equiv
    h(q)+\ldots+\frac{h^{(k-1)}(q)}{(k-1)!}(x-q)^{k-1}
\ee
Then the first term in (\ref{I}) is finite in the limit $q\to w$
and the second one can be integrated explicitly
\be
\int_w^\infty [h(q)]^{k-1}\frac{dx}{(x-q)^k\sqrt{x-w}}=
    \sum_{j=0}^{k-1} \frac{h^{(j)}(q)}{j!}
        \int_w^\infty \frac{dx}{(x-q)^{k-j}\sqrt{x-w}}=
\nonumber\\ \quad
    =\sum_{j=0}^{k-1} \frac{h^{(j)}(q)}{j!}
        \frac{\sqrt\pi}{(w-q)^{k-j-1}}\frac{\G(k-j+1/2)}{\G(k-j)}.
\ee

One can easily see that the divergent (as $q\to w$) part of $I$ exactly cancels the
residue part (\ref{Res}) and the summation formula in the limit
of $q=w$ takes the form
\be
\sum_{n=1}^\infty \frac1{(n^2+w^2)^{k+1/2}}
    - \int_0^\infty \frac{dx}{(x^2+w^2)^{k+1/2}}=
    \ee
    $$
    =-\frac{1}{2 w^{2 k+1}}
    +2(-1)^k \int_w^\infty \triangle h (x)
            \frac{dx}{(x-w)^{k+1/2}},
$$
where
\be
\triangle h (x) = h(x)-[h(x)]_w^{k-1}.
\ee

A more general case is also valid
$$
\sum_{n=1}^\infty \frac{\tilde f(n)}{(n^2+w^2)^{k+1/2}}
    - \int_0^\infty \frac{\tilde f(x) dx}{(x^2+w^2)^{k+1/2}}=
    $$
    $$
    =-i\pi\res{z=0}
        \(\frac{\tilde f(z)}{(z^2+w^2)^{k+1/2}}\frac{1}{1-e^{-2 \pi i n}}\)
    +2(-1)^k \int_w^\infty
            \frac{\triangle h (x)}{(x-w)^{k+1/2}}\,dx
$$
where $\tilde f(n)$ --- is a polynomial function of appropriate order
not to break (\ref{lim}), and
$$
    h(x)=\frac{\tilde f(x\, e^{i\pi/2})}{(x+w)^{k+1/2}}\,\frac1{e^{2\pi x}-1}
$$
Further generalizations of this formula for functions with two non-integrable
singularities and/or their combination with other know forms of APF is straightforward.

\section*{Acknowledgement}
We express sincere gratitude to Professor Aram Saharian for his kind help,
and to Dr Vladimir Markov for fruitful
discussions.

\end{document}